\documentclass{article}

\usepackage{authblk}
\usepackage{amsmath}
\usepackage{amssymb}
\usepackage{bbm}
\usepackage{amsthm}
\usepackage{hyperref}
\usepackage{mathtools}
\usepackage[a4paper]{geometry}
\usepackage{appendix}
\usepackage{mathtools}
\usepackage{tikz}
\usepackage{caption}
\usepackage{subcaption}
\usepackage{algorithm}
\usepackage{algpseudocode}
\usepackage{natbib}
\usepackage{graphicx}

\title{Clustering Multivariate Time Series using Energy Distance}

\author[1]{Richard A. Davis}
\author[1]{Leon Fernandes}
\author[2]{Konstantinos Fokianos}
\affil[1]{\small\textit{Department of Statistics, Columbia University}}
\affil[2]{\small \textit{Department of Mathematics and Statistics, University of Cyprus}}

\newcommand{\R}{\mathbb{R}}
\newcommand{\Z}{\mathbb{Z}}

\newcommand{\gid}{\ensuremath \xrightarrow{\hspace*{0.08cm} d \hspace*{0.08cm}}}

\newcommand{\gas}{\ensuremath \xrightarrow{a.s.}}
\newcommand{\E}{\mathbb{E}}
\newcommand{\RE}{\mathrm{Re}}
\newcommand{\Cov}{\mathrm{Cov}}
\newcommand{\eqd}{\stackrel{\mathclap{\tiny\mbox{$d$}}}{=}}
\newcommand{\innerproduct}[2]{\langle #1, #2 \rangle}

\numberwithin{equation}{section}

\newtheorem{theorem}{Theorem}[section]

\newtheorem{lemma}{Lemma}[section]

\newboolean{withgraph}
\setboolean{withgraph}{true}

\begin{document}

\maketitle

\begin{abstract}
A novel methodology is proposed for clustering  multivariate time series data using energy distance defined in Sz\'ekely and Rizzo (2013).
Specifically, a dissimilarity matrix is formed using the energy distance statistic to measure separation between the finite dimensional distributions for the component time series.
Once the pairwise dissimilarity matrix is calculated, a hierarchical clustering method is then applied to obtain the dendrogram.
This procedure is completely nonparametric as the dissimilarities between stationary distributions are directly calculated without making any model assumptions.
In order to justify this procedure, asymptotic properties of the energy distance estimates are derived for general stationary and ergodic time series.
The method is illustrated in a simulation study for various component time series that are either linear or nonlinear.
Finally the methodology is applied to two examples; one involves GDP of selected countries and the other is population size of various states in the U.S.A.~in the years 1900--1999.
\end{abstract}

\noindent
{\textbf{Keywords:} {Characteristic function; clustering; dissimilarity measure; energy distance; hierarchical clustering; stationarity; time series}}

\noindent
{\textbf{Mathematics Subject Classification:} Primary 62M10, 62H30; Secondary 62H20, 62H12.}

\section{Introduction}

Clustering is an important concept in statistics in which data is partitioned into groups where within each group, the data share similar characteristics.
By now, there are a plethora of clustering algorithms for forming such partitions (e.g., K-means and to some extent CART).
Most of these algorithms group data according to some notion of similarity (or dissimilarity) from which the data are then clustered into various groups; see \citet{XuTian2015} for a review of such procedures.
That is, the data are partitioned into the same group if each member is close to each other relative to a similarity measure.
The goal of this paper is to consider clustering of the component series in a multivariate time series setting, based on energy distance \citep[see][]{SzekelyRizzo2013} applied to the joint distributions of the component time series.
A key advantage of this method is that the procedure is nonparametric and based on measures of closeness of joint distributions as measured through their characteristic functions.
This is in contrast to parametric procedures where the clustering is performed via the parametric fitting to some family of models or to second order properties derived from autocorrelation functions; see references below.
\citet{Shumway1982} is an early and important contribution to discriminant analysis of time series that can be viewed as a precursor to the more general approach of our paper.
The characteristic function is always well defined even in  cases where we deal with multivariate (not necessarily normal) distributions.
Lemma \ref{lemma.denergy} shows that calculation of a suitable distance between characteristic functions is equivalent to calculation of the Euclidean distance between observations. 
Therefore, applying characteristic function techniques relieves the burden of distributional assumptions (which can be quite challenging in high dimensions)
and at the same time provides a computationally feasible way to implement multivariate time series clustering.  \citet{ZhangAn2018} considers clustering based on pairwise distributions via copulas.  However, due to the complexity of estimating joint distributions, their method was not extended to joint distributions beyond lagged pairs of observations.  

To fix ideas, consider a $d$-dimensional time series $\{\mathbf{X}_t = (X_{t1}, X_{t2}, \ldots, X_{td})^T,~t\in\Z\}$ whose component series are to be clustered.
Based on $n$ consecutive observations, say $\mathbf{X}_1, \mathbf{X}_2, \ldots,\mathbf{X}_n$, the prototypical strategy is to form a measure of dissimilarity between each pair of component series.
Once a measure of dissimilarity between the component series is decided upon, then a $d\times d$ dissimilarity matrix is computed.
This matrix is then used as the input to obtain the clustering via algorithms such as K-means, fuzzy C-means, spectral clustering and hierarchical clustering.
The true number of clusters, which is required for the former three methods, is typically not known \textit{a priori}; for this reason we shall focus on hierarchical clustering.
In this method each component at the initial step belongs to its own cluster; at each successive step, the most similar pairs of clusters are recursively merged.
The standard algorithms that facilitate cluster merging include complete linkage, single linkage, average linkage, centroid linkage and Ward's linkage \citep{Batagelj1988, Jamesetal2013, MurtaghLegendre2014}.
A hierarchy is obtained wherein the most similar components are in the same cluster and as one moves up the hierarchy, the clusters become more and more dissimilar.
The hierarchy is visualized as a dendrogram and is the main output of this algorithm.  Procedures such as the average silhouette width \citep[see][]{KaufmanRousseeuw2009} can be used to determine the final number of clusters from the dendrogram.

Various dissimilarity measures for clustering time series are catalogued in \citet{Liao2005, Fu2011, MonteroVilar2015, Aghabozorgi_etal2015,  MaharajD'UrsoCaido2019} and the references therein.
Typically one considers features such as autocorrelation, partial autocorrelation, periodogram, spectral density or the copula of joint distributions---the distances between these features forms the dissimilarity measure \citep[see][]{GaleanoPena2000, Caiadoetal2006, DiazVilar2010, ZhangAn2018}.
Dynamic time warping (DTW) utilizes a different approach where one finds an optimal mapping such that, the paired time series under the mapping minimizes a specific distance \citep[see][]{BerndtJames1994}.
Particularly relevant to our work is the paper of \citet{ZhangChen2018}, where a two dimensional version of the Kolmogorov-Smirnov statistic is used as the dissimilarity measure between distributions of lagged components.
The methods mentioned so far  are nonparametric, but with the exception of the copula procedure, are based primarily on second order properties of the processes.
As such the ``distance'' employed for clustering compares moments, not distributions, as it is developed later in this article.
Model based approaches typically assume the component time series are realizations of ARIMA or GARCH processes; clustering is then performed via the parametric fitting to the specified model.
Examples for distances in this class include Piccolo distance, Maharaj distance and cepstral-based distance \citep[see][]{Piccolo1990, Maharaj2000, Kalpakisetal2001, Savvidesetal2008}.
General dissimilarity measures between time series which are based on their spectral densities have been studied by \citet{Kakizawa_etall1998}, \citet[Ch. 6]{TaniguchiandKakizawa2000}.
Given two time series with spectral density matrices $f_{i}(.)$, $i=1,2$, these authors defined a dissimilarity (or disparity) measure by
$$D_{H}( f_{1}, f_{2}) = \frac{1}{4 \pi} \int_{-\pi}^{\pi} H(f_{2}^{-1}(\omega) f_{1}(\omega)) d \omega,$$
for a suitable function  $H(\cdot)$ which has to satisfy that $D_{H}( f_{1}, f_{2}) \geq 0$ and $ D_{H}( f_{1}, f_{2})= 0$ when $f_{1}(\omega)=f_{2}(\omega)$.
For instance, choosing $H(z)= z-\log (z)-1,  ~~z>0$ we have the Kullback-Leibler divergence.

In a time series setting, it is important to have dissimilarity measures that go beyond just the marginal distribution of the individual components.
That is, the dissimilarity measure should be based on the joint distributions of the individual lagged components.
Specifically, for a fixed lag $h \geq 0$, we consider the dissimilarity between the distributions of the $h+1$ dimensional time series $Y_t := (X_{t,j}, X_{t+1,j}, \ldots, X_{t+h,j})^T$ and $Z_t := (X_{t,k}, X_{t+1,k}, \ldots, X_{t+h,k})^T$ which will be measured through the energy distance of \citet{SzekelyRizzo2013}.

Energy distance, denoted by $d_E(Y_1,Z_1)$, see Section 2 for the definition, is nonnegative and has the property that $d_E(Y_1,Z_1)=0$ holds if and only if the joint distributions of $Y_t$ and $Z_t$ are the same.
This property for a dissimilarity measure holds only for the copula case.
We provide theoretical justification for the use of the energy distance statistic.
In particular, we derive asymptotic properties for the energy distance statistic for stationary ergodic time series.
Additionally, energy distance works well with heavy tailed data which is generally not the case for other dissimilarity measures.
Although we assume the components are of equal length, this is only for the sake of convenience; the results of this paper can be easily extended to the situation where the lengths of the time series vary across components.

The rest of the paper is organized as follows.
Section 2 defines the energy distance between any two distributions.
In Section 3 we present the main theorems on consistency and characterizing limit distributions of the energy distance statistic.
Section 4 discusses the multivariate time series clustering algorithm.
Various clustering tasks on simulated data are considered in Section 5; we experimentally obtain and compare the performance of the proposed methodology to some competing methods.
Our proposed procedure performed generally better when clustering nonlinear and multivariate VAR time series than other methods.
The ACF/PACF based procedures did well when the underlying component series are well differentiated by their second order properties such as, for example, Gaussian linear models.
In these situations,  some of the periodogram based methods  outperformed our method as did the ARMA-model based method.  However, this is not too surprising since these particular procedures are tuned well for this special class of models. Further details on the simulations and performance can be found in Section 5.
Two real world data sets are also considered.
The first is the annual GDP data for selected countries and the second involves the population growth for a number of states in the U.S.A.~in the years 1900--1999.
Proofs of all the main results in this paper are deferred to the Appendix.

\section{Energy Distance between Distributions}

Before introducing energy distance, it will be helpful to first fix some notation.
Throughout this paper, the inner-product between vectors $y,z\in\R^p$ is denoted by $\innerproduct{y}{z} = \sum_{j=1}^p y_jz_j$ and let $|y|^2 = \innerproduct{y}{y}$.
For a complex number $z=a+ib$, where $i$ is the imaginary number and the complex modulus is denoted by $|z| = \sqrt{a^2+b^2}$.
Further, if $X=(Y,Z)$ is a random vector then $\dot X = (\dot Y, \dot Z)$ and $\ddot X = (\ddot Y, \ddot Z)$ denote i.i.d copies of $X$.

Let $Y$ and $Z$ denote $p$-dimensional random vectors with characteristic functions $\varphi_Y$ and $\varphi_Z$ respectively.
The energy distance \citep{SzekelyRizzo2013} between $Y$ and $Z$ is defined by
\begin{align}
	\label{eq_denergy}
	d_E(Y,Z) := \int_{\R^p} \big| \varphi_Y(s) - \varphi_Z(s) \big|^2 d\mu(s),
\end{align}
where $\mu$ is the infinite measure given by
\begin{align}
	\label{eq.mu}
	d\mu(s) = \frac{ds}{|s|^{p+1}c_p},
\end{align}
and $c_p = \pi^{(p+1)/2}/\Gamma\big((p+1)/2\big)$.
Other weight functions can also be used; see Remark 1 for more details.
The significance of using this particular weight function in evaluating \eqref{eq.mu} is that the integral \eqref{eq_denergy} can be explicitly calculated, as shown in the following lemma.
\begin{lemma}
\label{lemma.denergy}
Consider random vectors $Y,Z$ in $\R^p$.
If $\E[ |Y| + |Z|] < \infty$ then $d_E(Y,Z) < \infty$.
Furthermore,
\begin{align}
	\label{eq_denergy.exp}
	d_E(Y,Z) = 2\E|Y-\dot Z| - \E|Y-\dot Y| - \E|Z-\dot Z|.
\end{align}
\end{lemma}
The proof is provided in the Appendix.
Note that \eqref{eq_denergy.exp} provides a simple formula for $d_E(Y,Z)$ and it only depends on Euclidean distances between random vectors.
Lemma \ref{lemma.denergy} implies calculation of $d_E(Y,Z)$ requires finite first moments to guarantee finiteness of the energy distance statistic.
The expression we obtain here is similar to the maximum mean discrepancy (MMD), as given by \citet{Gretton_etal2012}.

It is clear from \eqref{eq_denergy} that $d_E(Y,Z)$ is nonnegative and equality holds if and only if $Y$ and $Z$ have the same distribution.
As mentioned earlier, this simple observation is the basis for the clustering methodology that we consider in the following section.
Indeed we can obtain sample estimates for $d_E(Y,Z)$ and then calculate a dissimilarity measure between the distributions of $Y$ and $Z$; consequently, $d_E(Y,Z)$ can be employed for clustering.
The method is completely nonparametric and easy to implement.
For illustration purposes we display the theoretically calculated energy distance in the two examples below.

\paragraph{Example 1.}
Let $Z\sim \mathcal{N}(0,1)$ have a standard normal distribution.
For $t\in\R$ a straightforward calculation yields $\E|Z-t| = |t|(2\Phi(|t|)-1) + \sqrt{\frac{2}{\pi}} e^{-t^2/2}$, where $\Phi(\cdot)$ is the cdf of the standard normal.
It then follows that for any $\theta \in\R$ and $\sigma>0$,
\begin{align*}
	d_E (\sigma Z+\theta, Z)&= 2\sqrt{\sigma^2+1} \left( |\tau| \Big( 2 \Phi \big(|\tau|\big)  - 1\Big) + \sqrt{\frac{2}{\pi}}e^{-\tau^2/2} \right) - \frac{2}{\sqrt \pi} (\sigma+1),
\end{align*}
where $\tau = \theta/\sqrt{\sigma^2+1}$.

\paragraph{Example 2.}
Let $Y \sim \mathcal{L}(0,\lambda)$ have a Laplace distribution with density $f_Y(y) =  e^{-|y|/\lambda}/(2\lambda)$, where $\lambda > 0$.
For $t\in\R$, $\E|Y-t| = \lambda e^{- |t|/\lambda} + |t|$.
Then,
\begin{align*}
	d_E(Y, Z) &= 4 \lambda \big( 1-\Phi(\lambda^{-1}) \big) \exp \Big( \frac{1}{2\lambda^2} \Big) - \frac{3\lambda}{2} + \frac{2(\sqrt{2}-1)}{\sqrt{\pi}}.
\end{align*}

\paragraph{Remark 1.}
Alternatively, consider probability measures instead of $\mu$ in \eqref{eq.mu}, such as a Gaussian measure; see \citet{HongGuWhitehouse2017}.
Employing a probability measure guarantees finiteness of the integral without assuming that the random vectors $Y$ and $Z$ have finite means.
 Although we use the $\mu$ in \eqref{eq.mu} associated with energy distance, similar results can be obtained with $\mu$ replaced by a probability measure. 
We give a few details to be more specific: let $d_{P}(Y,Z)$ denote the distance \eqref{eq_denergy} with $\mu$ replaced by a probability measure $\mu_P$.
Then, easy calculations show that the counterpart of \eqref{eq_denergy.exp} is given by
\begin{align*}
d_{P}(Y,Z) = \E \RE \varphi_{P}(Y-\dot Y) + \E \RE \varphi_{P}(Z-\dot Z) - 2 \E \RE \varphi_{P}(Y-\dot Z),
\end{align*}
where $\varphi_{P}$ is the characteristic function of $\mu_P$ and $\RE(\cdot)$ denotes the real part of a complex number.
Hence, choosing $\mu_P$ whose characteristic function is explicitly known yields different formulas for $d_P(Y,Z)$.
For example if $\mu_P$ is the Gaussian measure given by $d\mu_P(s) =  (\sqrt{2\pi\sigma^2})^{-1} \exp \big(-s^2/(2\sigma^2) \big) ds$ for $s\in\R$ and some $\sigma^2>0$, then $\RE \varphi_{P}(s) = \exp(-\sigma^2s^2/2)$ so that
\begin{align*}
	d_{P}(Y,Z) = \E [ \exp(-\sigma^2(Y-\dot Y)^2/2 ] + \E [\exp(-\sigma^2(Z-\dot Z)^2/2] - 2 \E [\exp(-\sigma^2(Y-\dot Z)^2/2].
\end{align*}

\section{Empirical Energy Distance Statistic for Time Series}
Let $\{(Y_t, Z_t)\}$ be a stationary and ergodic time series, where $Y_t,Z_t\in\R^p$.
Denote the stationary distribution of this process by $(Y,Z)$.
We will now show how to empirically estimate $d_E(Y,Z)$ based on observations $(Y_1,Z_1), \ldots,(Y_n,Z_n)$ using the empirical characteristic function and obtain the asymptotic properties of this estimator.  Although we have assumed that the sample sizes of $\{Y_t\}$ and $\{Z_t\}$ are the same, this is not necessary since we are only interested estimating the marginal characteristic functions.  It is straightforward to adapt our results to the case of unequal sample sizes.
Let $\hat\varphi_Y(s) := \frac1n\sum_{j=1}^n e^{i \innerproduct{s}{Y_j}}$ and similarly define $\hat\varphi_Z(s)$.
The estimate of $d_E(Y,Z)$ is given by
\begin{align*}
	\hat d_E(Y,Z) := \int_{\R^p} \big| \hat \varphi_Y(s) - \hat \varphi_Z(s) \big|^2 d\mu(s).
\end{align*}
Leveraging \eqref{eq_denergy.exp} we can write $\hat d_E(Y,Z)$ as the $V$-statistic,
\begin{align}
	\label{eq.hat.V}
	\hat d_E(Y,Z) = \frac{2}{n^2} \sum_{j,k=1}^{n} |Y_j-Z_k| -  \frac{1}{n^2} \sum_{j,k=1}^{n} |Y_j-Y_k| -  \frac{1}{n^2} \sum_{j,k=1}^{n}|Z_j-Z_k|.
\end{align}
We thus have the sample estimate $\hat d_E(Y,Z)$ that can be computed easily and it is based solely on the distance between the observations.
Note that \eqref{eq.hat.V} is computable even in the case of multivariate observations (dependent or not).
The following theorem shows that $\hat d_E(Y,Z)$ is a consistent estimator for $d_E(Y,Z)$:
\begin{theorem}
\label{thm_cons_direct}
Consider stationary and ergodic time series $\{(Y_t,Z_t)\}$, where $Y_t,Z_t\in\R^p$ and let $(Y,Z)$ have the same distribution as $(Y_1,Z_1)$.
Assuming $\E[ |Y| + |Z| ] < \infty$, we have as $n\rightarrow\infty$
\begin{align}
	\label{cons.gen}
	\hat d_E(Y,Z) \gas d_E(Y,Z).
\end{align}
\end{theorem}
The proof is provided in the Appendix
and it involves studying the asymptotic behavior of the empirical characteristic function process.
To obtain the asymptotic distribution, we need additional moment assumptions and the notion of weak dependence.
In what follows, assume that $\{(Y_t,Z_t)\}$ is an $\alpha$-mixing time series with rate function $\alpha(h)$.
Recall the definition of $\alpha$-mixing  \citep[][p.~18]{Doukhan1994}: for integers $h\ge 0$,  the $\alpha$ mixing rate function is defined by
\begin{equation*}
	\alpha(h)=\sup|\mathbb{P}(U\cap V)-\mathbb{P}(U)\mathbb{P}(V)|\,,
\end{equation*}
where the suprema is taken over $U\in\sigma((Y_s,Z_s),~s=\ldots,-1,0)$, and $V\in \sigma((Y_s,Z_s),~s=h+1,h+2,\ldots)$, respectively.
The process is then said to be $\alpha$-mixing if $\alpha(h)\to0$ as $h\to\infty$.
The following theorem characterizes the asymptotic distributions,
whereby the rate of convergence differs according to whether or not the distributions of $Y$ and $Z$ are equal.
\begin{theorem}
\label{thm_asymptotic_direct}
Consider stationary and ergodic time series $\{(Y_t,Z_t)\}$ where $Y_t,Z_t\in\R^p$ such that $\sum_h {\alpha(h)}^{1/r} < \infty$ for some $r>1$.
Set $u = 2r/(r-1)$ and write $Y_1= (Y_{11},\ldots,Y_{1,p})^T$ and $Z_1 = (Z_{11},\ldots,Z_{1p})^T$.
Assume that for some $\alpha \in (u/2,u]$ the following hold:
\begin{align}
\label{eq.same.moment}
\E[ |Y_1|^{\alpha} + |Z_1|^{\alpha} ] < \infty \text{ and }  \E \bigg[ \bigg(1 \vee \prod_{\ell=1}^p |Y_{1\ell}|^{\alpha}\bigg) \bigg(1 \vee \prod_{\ell=1}^p |Z_{1\ell}|^{\alpha}\bigg) \bigg] < \infty.
\end{align}
\begin{enumerate}
\item
If $Y_1$ and $Z_1$ have the same distribution then,
\begin{align*}
	n\hat d_E(Y,Z) \gid ||G||^2_{\mu} = \int_{\R^p} |G(s)|^2 d\mu(s),
\end{align*}
where $G(s)$ is a complex-valued mean-zero Gaussian process with covariance structure for $s,t\in\R^p$ given by
\begin{align}
	\label{eq.cov}
	\Cov(G(s),G(t)) &= \sum_{h \in \mathbb{Z}} \Cov \big( e^{i \innerproduct{s}{Y_0}} - e^{i \innerproduct{s}{Z_0} } , e^{i \innerproduct{t}{Y_h}} - e^{i \innerproduct{t}{Z_h} } \big).
\end{align}
\item
If $Y_1$ and $Z_1$ do not have the same distribution then,
\begin{align*}
	\sqrt{n} ( \hat d_E(Y,Z) - d_E(Y,Z)) \gid G'_{\mu} = \int_{\R^p} G'(s) d\mu(s),
\end{align*}
where $G'(s) = 2\RE[(\varphi_Y(s)-\varphi_Z(s))\cdot \overline{G(s)}]$.
\end{enumerate}
\end{theorem}

Theorems \ref{thm_cons_direct} and \ref{thm_asymptotic_direct} state that under minimal assumptions $\hat d_E(Y,Z)$ is a consistent estimator for $d_E(Y,Z)$ and under additional moment and mixing conditions, converges in distribution, suitably normalized.
The rates of convergence are different depending on whether or not $Y$ is equal in distribution to $Z$.  In particular, we see that $n\hat d_E(Y,Z)$ converges to a non-degenerate random variable when $Y$ and $Z$ have the same distribution, but tends to infinity otherwise.

\section{Multivariate Time Series Clustering}

\subsection{Dissimilarity metric based on $(h+1)$-dimensional joint distributions}

Consider observations $\{\mathbf{X}_1, \mathbf{X}_2, \ldots, \mathbf{X}_n\}$ from a multivariate time series $\mathbf{X}_t = (X_{t1}, X_{t2}, \ldots, X_{td})^T$ in $\R^d$.
In this section, a general methodology for clustering component time series based on $(h+1)$-dimensional distributions, for a fixed lag $h \geq 0$, is presented.
We compute a pairwise dissimilarity matrix using the energy distance on these joint distributions and then apply a hierarchical clustering algorithm to classify the data.

The dissimilarity measure based on the $h$-lagged $j^{th}$ and $k^{th}$ components of $\mathbf{X}_t$ is given by $D_{jk} = \hat d_E(Y,Z)$, where $Y_t  = (X_{t,j}, X_{t+1,j}, \ldots, X_{t+h,j})^T$ and $Z_t  = (X_{t,k}, X_{t+1,k}, \ldots, X_{t+h,k})^T$ for $t=1,\ldots,n-h.$ The energy distance dissimilarity measure is given by $ D_{jk} := \hat d_E(Y,Z)$; see also \citet{FokianosPitsillou2018}.
Note that $D_{jj} = 0$ and due to symmetry, $D_{kj} = D_{jk}$.
In this way, we form the energy distance dissimilarity matrix $D = [D_{jk}]_{j,k=1}^p$.

To obtain the clustering, an agglomerative hierarchical clustering method \citep[see for example,][]{Batagelj1988} is used which we briefly describe below.
We start with the original $d$ components, as nodes, and successively merge nodes (or clusters) to form new clusters.
The inter-cluster dissimilarities are then obtained as $d(C_j,C_k) = D_{jk}$ for $1\leq j\neq k \leq d$.
The least dissimilar pair of components, say $C_j$ and $C_k$, are now merged; note that only $d-1$ inter-cluster dissimilarities need to be updated.
We employ the generalized Ward's linkage algorithm in this paper which has an update formula for computing the dissimilarity between the merged cluster $C_j\cup C_k$ with $C_\ell$ for $\ell \ne j,k$.
This formula, known as the Lance-Williams formula for generalized Ward's linkage, is given by
\begin{equation}
	\label{eq.lance}
	d(C_j \cup C_k, C_{\ell}) = \frac{n_j+n_{\ell}}{n_j+n_k+n_{\ell}} d(C_j, C_{\ell}) + \frac{n_k+n_{\ell}}{n_j+n_k+n_{\ell}} d(C_k, C_{\ell}) - \frac{n_{\ell}}{n_j+n_k+n_{\ell}} d(C_j, C_k),
\end{equation}
where $n_i$ denotes the number of components in cluster $C_i$.
Proceeding forward, suppose now there are $J$ clusters $C_1,\ldots,C_J$ with a $J\times J$ inter-cluster dissimilarity matrix $D$.  The pair $C_j, C_k$ with the least dissimilarity  are  merged with the resulting
 inter-cluster dissimilarities (for the $J-1$ clusters) are given by \eqref{eq.lance}.
The clustering algorithm is summarized in Algorithm \ref{alg:declust}.

\begin{algorithm}[H]
\caption{Time Series Clustering using Energy Distance}\label{alg:declust}
\begin{algorithmic}
\item\textbf{Input:} $\{\mathbf{X}_1, \mathbf{X}_2, \ldots, \mathbf{X}_n\}$, $h$
\For{$j<k$ where $j,k\in\{0,1,\ldots,d\}$}
\State $Y_t \gets (X_{t,j}, X_{t+1,j}, \ldots, X_{t+h,j})$
\State $Z_t \gets (X_{t,k}, X_{t+1,k}, \ldots, X_{t+h,k})$
\State $D_{jk} \gets \hat d_E(Y,Z)$
\EndFor
\State Initialize clusters $C_1, \ldots,C_d$ where $C_j$ contains the $j^{th}$ component time series.
\State Set the inter cluster dissimilarities $d(C_j, C_k) \gets D_{jk}$, $1\leq j\neq k \leq d$.
\For{$J = d, d-1, \ldots, 2$}
\State Identify $1\leq j \neq k \leq J$ with smallest $d(C_j,C_k)$. Merge to form $C_j \cup C_k$.
\State Update the inter cluster dissimilarities for $J-1$ clusters using \eqref{eq.lance}.
\EndFor
\end{algorithmic}
\end{algorithm}

If the true number of clusters ($K_0$) is known, then one can obtain $K_0$ clusters from the hierarchical clustering.
As the true number of clusters are usually not known, metrics such as the average silhouette width \citep{KaufmanRousseeuw2009} can be used to determine the final number of clusters.
The silhouette coefficients are defined for each component and are based on the tightness and separation of the clusters using the dissimilarity matrix $D$.
Specifically, if we obtain $K$ clusters $\{C_1, \ldots,C_K\}$ from the hierarchical clustering, then the $i^{th}$-node silhouette coefficient  is defined as
\begin{align*}
	s(i) = \frac{d(i, C_k) - d(i, C_j)}{\max ({d(i, C_k), d(i, C_j))}},
\end{align*}
where $d(i, C_{\ell})$ is the average dissimilarity of component $i$ to all components in $C_{\ell}, 1\leq \ell \leq K$, $C_j$ is the cluster that contains the $i^{th}$ component and $C_k$ is the cluster which satisfies $d(i, C_k) = \min_{\ell \neq j} d(i, C_{\ell})$.
Clearly $-1 \leq s(i) \leq 1$ and the closer $s(i)$ is to one, the better the quality of the clustering.
The average silhouette width is obtained as the average value of $s(i)$ among all the $d$ components.
The average silhouette width is computed for $K$ clusters for $2\leq K < d$.
The value of $K$ which maximizes the average silhouette width is the most appropriate choice for the number of clusters.

\subsection{Clustering via Lagged Bivariate Distributions}

Instead of using $(h+1)$-dimensional distributions for comparison one could simplify and restrict attention to lagged bivariate distributions.
More precisely, for each $\ell \in\{1, \ldots,h\}$ compute the energy distance $D_{j,k}^{(\ell)} = \hat d_E(Y,Z)$, where $Y_t=(X_{t,j}, X_{t+\ell,j})$ and $Z_t=(X_{t,k}, X_{t+\ell,k})$.
We also include $D^{(0)}$, the lag zero dissimilarity matrix which computes the pairwise dissimilarity between the marginal distributions of the components.
This yields $(h+1)$ dissimilarity matrices, $D^{(\ell)}$ for $\ell = 0,\ldots, h$
which contain the distances between $X_t$ and $X_{t+\ell}$.
An overall (total) dissimilarity matrix is defined by $D = \sum_{\ell=0}^h D^{(\ell)}$.
A similar approach can be found in \citet{ZhangAn2018} and \citet{ZhangChen2018} where a (weighted) sum of dissimilarity matrices up to some maximum lag is used as the final dissimilarity matrix.
The hierarchical clustering method described Subsection 4.1 is then applied to $D$.

\section{Empirical Comparisons and Applications}

To assess the performance of our method we consider several simulated data sets.
In addition, we apply our method to two real data sets in this section.
Three clustering tasks with simulated data are considered, where for the first two, the experiments of \citet{DiazVilar2010} are performed for comparison.
Next, the third simulated data set consists of clustering the components of a 40 dimensional VAR time series.
The two real data sets concern the G.D.P data of the world's most developed countries between, as observed between 1990 to 2011, and the populations of a subset of states in the U.S.A.~between the years of 1900 to 1999.

\subsection{Simulation Examples}

In the experiments we applied Algorithm \ref{alg:declust} with different lags $h=0,1,2$ and $5$.  See Remark 2 below for comments on the choice of lag $h$.
For comparison, various competing clustering methods using different dissimilarity measures (15 in total) that have been proposed in the existing literature were considered.
In the ACF based methods \citep[see][]{GaleanoPena2000}, the dissimilarity measure is equal to a geometrically downweighted distance between the estimated ACF of each pair of components.
In symbols, if $\hat\rho_{\ell,X_j}$ and $\hat\rho_{\ell,X_k}$ represent the estimated autocorrelations of the $j^{th}$ and $k^{th}$ components at lag $\ell$ respectively, then the dissimilarity measure is given by
$\big(\sum_{\ell=1}^L p(1-p)^{\ell} \cdot(\hat\rho_{\ell,X_j}-\hat\rho_{\ell,X_k})^2\big)^{1/2}$, where $L$ is the maximum lag considered and $0<p<1$.
In our simulations we take $p=0.05$ and $L=10,25,50$.
The analogous PACF based methods are also considered wherein the estimated ACF coefficients are replaced with the corresponding estimated PACF coefficients.  In the graphs displaying the results below, these procedures are labeled by {\it ACFLh} and {\it PACFLh} while the energy-based methods are labeled {\it EnergyLh}, where $h$ is the lag.

We also compared these methods with periodogram-based methods of \citet{Caiadoetal2006} that calculates Euclidean distances between the periodograms and log-periodograms.
In addition, integrated periodogram of \citet{Casado2010}, which computes the integral difference between the cumulative versions of the periodograms, is also included.
Finally, the ARMA model based dissimilarity measures developed in \citet{Piccolo1990} and \citet{Maharaj2000} were also compared.
The above methods obtain different dissimilarity matrices from which hierarchical clustering is then performed using generalized Ward's method.
We found that in our simulations, the clustering performance with energy distance was the highest with Ward's linkage.
Other linkage algorithms with respect to competing dissimilarity measures did not have significantly different performance.
These methods are labeled as {\it PER}, {\it PER.LP}, {\it INT.PER}, {\it AR.MAH}, and {\it AR.PIC  } in the graphs below. 

When the ground truth is known, we can compare the clustering methods using clustering evaluation metrics.
Specifically, we consider the similarity index of  \citet{Gavrilovetal2000} which is defined as
\begin{equation*}
	\mathit{Sim}(G, A) = \frac{1}{K} \sum_{i=1}^K \max_{1\leq j \leq K} \mathit{Sim}(G_i, A_j),
\end{equation*}
where $G=\{G_1, \ldots,G_K\}$ is the ground truth of the $K$ clusters, $A=\{A_1, \ldots,A_K\}$ is the clustering to be assessed, and
\begin{equation*}
	\mathit{Sim}(G_i, A_j) = \frac{2|G_i \cap A_j|}{|G_i|+|A_j|}.
\end{equation*}
Here $|\cdot|$ denotes the cardinality of a set.
The similarity index takes values between 0 and 1, with 1 corresponding to perfect clustering, that is, $G$ and $A$ are identical.
Other metrics were also considered such as the Rand index \citep{Rand1971}, adjusted Rand index \citep{HubertArabie1985} and a leave-one-out cross-validation \citep{Tanetal2006}.
However, the choice of metric did not have much impact on the relative comparisons between the various methods and we will only report the similarity index in our results.

\ifthenelse{\boolean{withgraph}}
{
\begin{figure}[t]
\centering
\scalebox{0.8}{\input{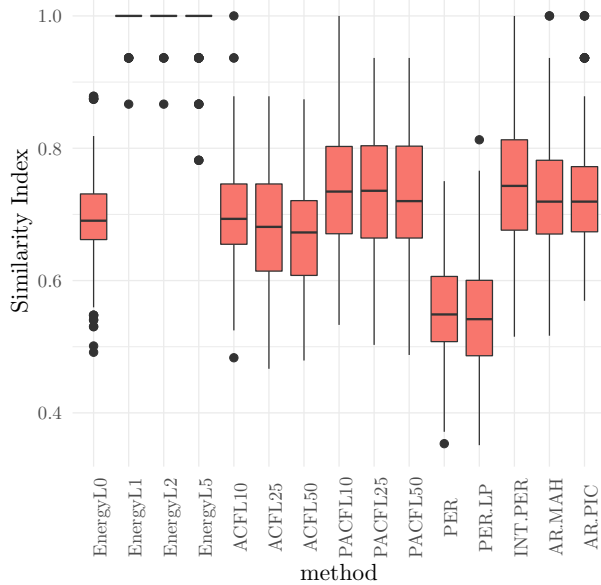}}
\caption{Comparison of similarity index of clustering methods of multivariate time series consisting of nonlinear components generated from Example 5.1. Results are based on 200 simulations and $n=200$.}
\label{fig_non_linear}
\end{figure}
}

\paragraph{Example 5.1}
For the first experiment, a time series consisting of 16 independent components is generated.
Each component is of length $n=200$ and is rescaled to have mean zero and standard deviation one.
We consider four clusters each of which contain four time series generated by the following models:
(i) threshold autoregressive (TAR) $X_t = 0.5 X_{t-1} I(X_{t-1} \leq 0)  - 2 X_{t-1} I(X_{t-1} > 0) + \varepsilon_t$,
(ii) exponential autoregressive (EXPAR) $X_t = (0.3 -10\exp (-X_{t-1}^2 ) )X_{t-1} + \varepsilon_t$,
(iii) linear moving average (MA) $X_t = \varepsilon_t - 0.4\varepsilon_{t-1}$ and
(iv) nonlinear moving average (NLMA) $X_t = \varepsilon_t - 0.5\varepsilon_{t-1} + 0.8\varepsilon^2_{t-1}$.
The sequence $\{\varepsilon_t\}$ is assumed to be iid with a standard normally distribution in all cases.
The similarity index is obtained against the ground truth of $K_0=4$ clusters, for each of the 15 methods.
This experiment was repeated 200 times and the boxplots of the similarity index are shown in Figure \ref{fig_non_linear}.
The dots in the figure denote the outliers in the boxplots.
In particular, our methods with positive lags have perfect clustering in all but three instances in our experiments.
It is clear that the energy distance based methods, except for lag 0, are nearly perfect in recovering the true clusters.

\ifthenelse{\boolean{withgraph}}
{
\begin{figure}[t]
\centering
\scalebox{0.8}{\input{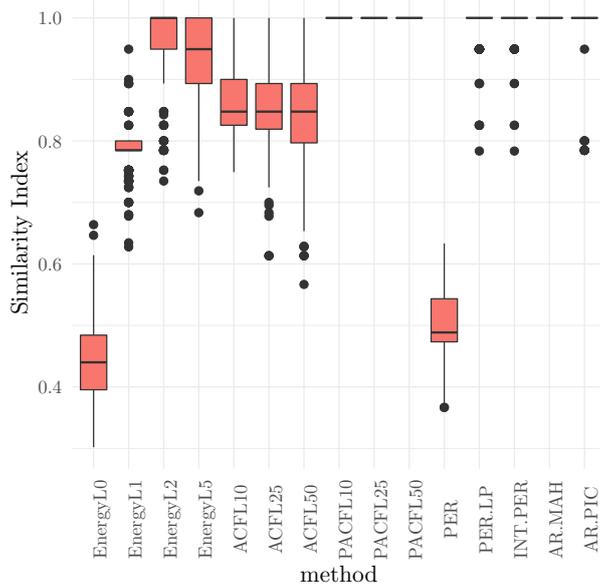}}
\caption{Comparison of similarity index of clustering methods of multivariate time series consisting of linear components generated from Example 5.2. Results are based on 200 simulations and $n=1000$.}
\label{fig_arma}
\end{figure}
}

\paragraph{Example 5.2}
A very similar setup is considered in this example where we instead have five clusters, with four time series each, from the following ARMA models:
(i) AR(1): $X_t = 0.5 X_{t-1} + \varepsilon_t$,
(ii) MA(1): $X_t = 0.7 \varepsilon_{t-1} + \varepsilon_t$,
(iii) AR(2): $X_t = 0.6 X_{t-1} + 0.2 X_{t-2} + \varepsilon_t$,
(iv) MA(2): $X_t = 0.8 \varepsilon_{t-1} - 0.6 \varepsilon_{t-2} + \varepsilon_t$,
(v) ARMA(1,1): $X_t = 0.8 X_{t-1} + \varepsilon_t + 0.2 \varepsilon_{t-1}$.
The sequence $\{\varepsilon_t\}$ is iid normally distributed.
In this simulation we consider the lengths of the time series to be $n=1000$.
From Figure \ref{fig_arma} we see that the best performance is achieved by $d_{AR.MAH}$, $d_{AR.PIC}$ and the PACF based methods, followed by $d_{INT.PER}$ and $d_{PER.LP}$.
This is due to the ARMA coefficients, PACF and log-periodograms being separated.
It is not surprising that $d_E$ with lag 0 performed the worst because all the marginal distributions were standard normal in the case.
The ACF based methods performed worse because the theoretical auto-covariance coefficients are not well separated.
Indeed, the true ACFs of the AR(2) and ARMA(2) considered here are very close to each other.
Furthermore, the autocorrelation at lag 1 of the AR(1) is 0.5 and that of MA(1) is 0.47; even though the true ACF of the MA(1) process is exactly zero for higher lags while that of the AR(1) decreases geometrically by a factor of 0.5, in practice this means that the estimated ACFs of the MA(1) and AR(1) will be rather close as well.
As the ACF determines the joint distributions for Gaussian ARMA time series, the energy based methods do see a reduction in performance.
However, with the correct specification of the lag $L=2$, we observe that the energy distance find the correct clustering most of the time.

\ifthenelse{\boolean{withgraph}}
{
\begin{figure}[t]
\centering
\scalebox{0.8}{\input{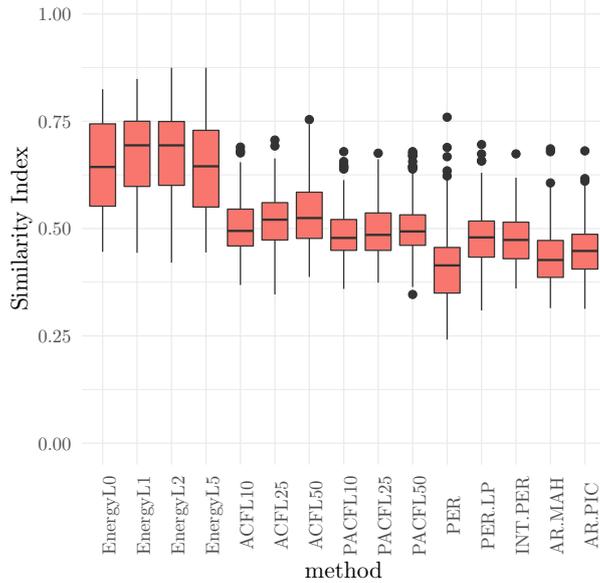}}
\caption{Comparison of similarity index of clustering methods of multivariate time series consisting of components generated from Example 5.3. Results are based on 200 simulations and $n=200$.}
\label{fig_multi}
\end{figure}
}

\paragraph{Example 5.3}
In this example, we consider clustering a 40 dimensional time series.
We generate four independent multivariate time series, each belonging to $\R^{10}$ according to the following models.
(i) VAR(1), $\mathcal{N}$: $X_t = B X_{t-1} + \varepsilon_t$, where the $10 \times 10 $ matrix $B$ is constructed as $100$ equally spaced numbers between $-1$ and $1$ column-wise which is then standardized to have spectral norm less than $1$.
The sequence $\{\varepsilon_t\}$ is iid $\mathcal{N}_{10}(\mathbf{0},I_{10})$.
(ii) VAR(1), $t_2$: $X_t = B X_{t-1} + \varepsilon_t$ where $B$ is the same as above and the only change being that the components of $\varepsilon_1$ are independent with a Student's t distribution with 2 degrees of freedom.
(iii) VAR(2), $\mathcal{N}$: $X_t = B_1 X_{t-1} + B_2 X_{t-2} + \varepsilon_t$, where similar to $B$, the $10 \times 10 $ matrices $B_1$ and $B_2$ are constructed using $100$ equally spaced numbers between $-1$ and $0$, and $0$ and $1$ respectively.
$B_1$ and $B_2$ are standardized using the maximum eigenvalue of $(B_1+B_2)(B_1+B_2)^T$.
The $\{\varepsilon_t\}$ is iid $\mathcal{N}_{10}(\mathbf{0},I_{10})$.
(iv) VAR(2), $t_2$: $X_t = B_1 X_{t-1} + B_2 X_{t-2} + \varepsilon_t$ with the only change being that the components of $\varepsilon_1$ are independent and distributed as Student's t with 2 degrees of freedom.
With these four underlying clusters, the clustering performance with respect to similarity index is shown in Figure \ref{fig_multi}.
Our proposed method outperformed all the competing methods.
The clustering performance was nearly the same for the energy distance based method for lags $h=0,1,2,5$.
This suggests that components with the same marginal distributions have been clustered correctly and inclusion of further lagged joint distributions did not appreciably improve clustering performance.

\paragraph{Remark 2.} In applications we need to decide on a suitable choice for $h$, the size of the joint distributions used for clustering.
If there is clustering at lag $h=0$, then one would expect clustering to also be present at lags $h>0$.
However, the nature of the clustering could be different as a function of lag.
For instance there might be strong associations or disassociation in the component time series at lag 3 which is not so evident at lags 0--2.
The other hurdle is that as the lag increases, the dissimilarity measure may incur more noise and hence less useful for providing meaningful clusters.
On the other hand, for $h$ small, there may not be much power in  discriminating the individual time series.
To some extent, this sort of behavior is manifested in Figures \ref{fig_arma} and \ref{fig_multi}.
In the case $h=0$ our proposed clustering procedure does a reasonably good job in correctly identifying the clusters.
This performance is only improved as one uses $h=1$ and $h=2$.
However, for $h=5$ there is a slight falloff in the performance of the clustering.
The idea of choosing an {\it optimal} $h$ will be the subject of a future investigation.

\subsection{Real Data Examples}

\ifthenelse{\boolean{withgraph}}
{
\begin{figure}[t]
\centering
\resizebox{.9\linewidth}{!}{\input{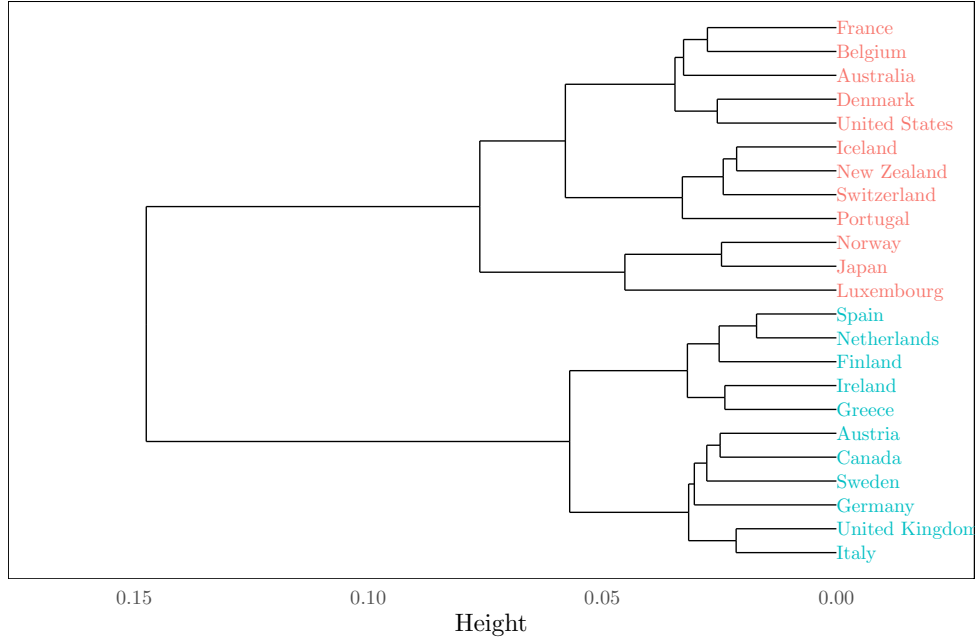}}
\caption{Clustering dendrogram obtained from real GDP data with energy distance using lag 1.
The colors correspond to the final clustering based the highest average silhouette width of 0.29 obtained with 2 clusters.}
\label{fig_new_gdp}
\end{figure}
}

\subsubsection*{Application to annual G.D.P. data}

In this example, we considered the annual real gross domestic product (GDP) data obtained from \href{https://www.conference-board.org/us/}{https://www.conference-board.org/us/}, which was also studied in \citet{ZhangAn2018}.
A set of 23 of the most developed countries in the world was used for the years 1980--2019:
Austria, Belgium, Denmark, Finland, France, Germany, Greece, Iceland, Ireland, Italy, Luxembourg, the Netherlands, Norway, Portugal, Spain, Sweden, Switzerland, United Kingdom, Canada, United States, Australia, New Zealand and Japan.
In total this is a 23 dimensional time series with $n=39$ observations each.
We used the annual log growth rate, calculated for the $j^{th}$ country as $\log (GDP_{t,j}) - \log (GDP_{t-1,j})$ as the input time series.
Each components was re-normalized to have mean zero and standard deviation one.
Energy distance with lag 1 is applied to this dataset; higher values of lag did not work well due to $n=39$ being too small.

In this case, the maximum average silhouette width was obtained with 2 clusters and is shown in Figure \ref{fig_new_gdp}.
Figure \ref{fig_gdp_zoom} presents the the clustering on a world map.
Spain, Netherlands, Finland, Ireland, Greece, Austria, Canada, Sweden, Germany, United Kingdom and Italy form the blue cluster.
This cluster consists of most mainland European countries and includes the United Kingdom, Ireland and Canada.
The red cluster consisted of France, Belgium, Australia, Denmark, United States, Ireland, Iceland, New Zealand, Switzerland, Portugal, Norway, Japan and Luxembourg.
A possible interpretation of this result is that the blue group includes countries that spend a considerable amount of their budget on safety net programs when compared to most of the countries in the red cluster.

\ifthenelse{\boolean{withgraph}}
{
\begin{figure}[t]
\centering
\includegraphics[width=0.9\linewidth]{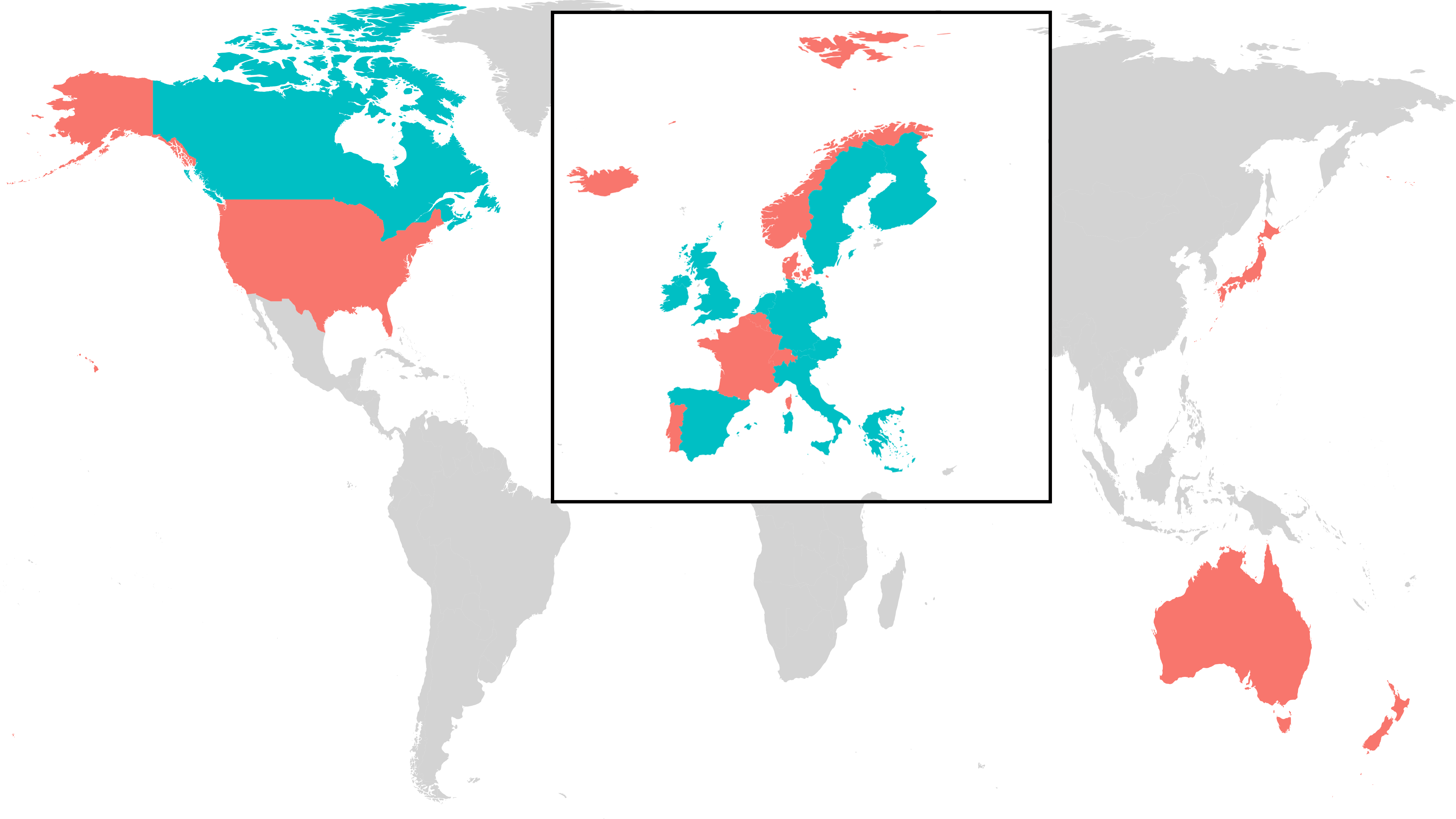}
\caption{Clustering obtained from real GDP data using energy distance with lag 1 on a world map with a zoomed in Europe.}
\label{fig_gdp_zoom}
\end{figure}
}

\subsubsection*{U.S.A.~Population Data}

\ifthenelse{\boolean{withgraph}}
{
\begin{figure}[t]
\centering
\input{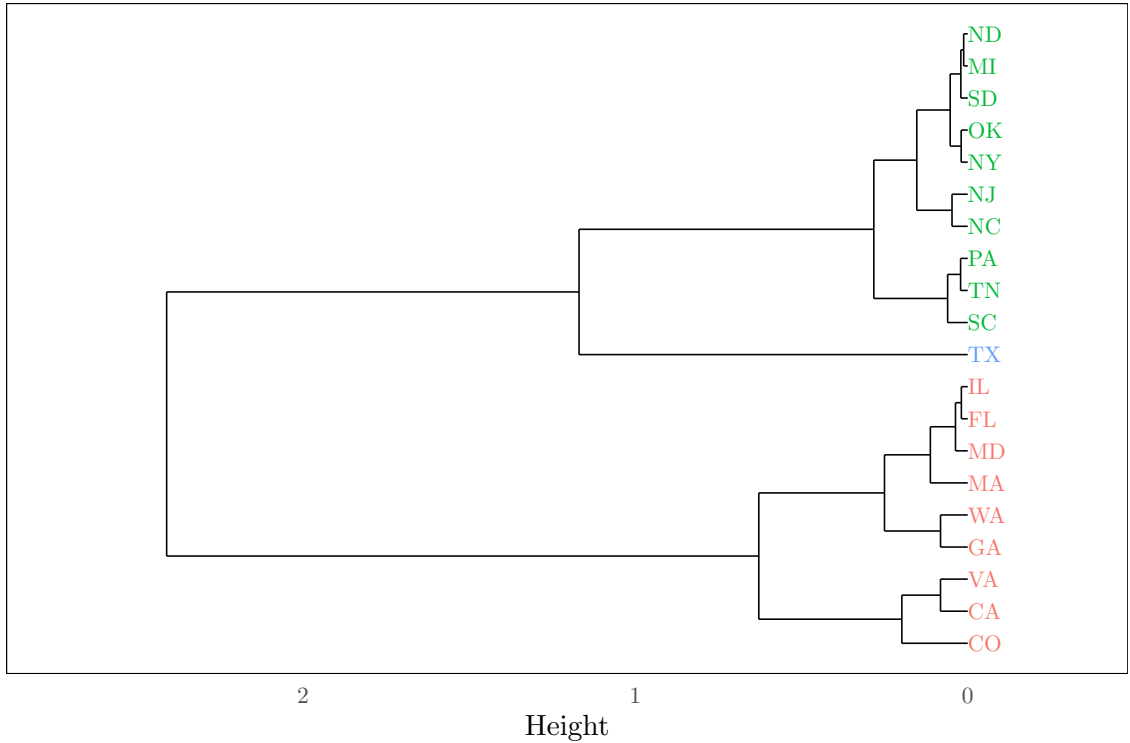}
\caption{Clustering dendrogram obtained from population data using energy distance with lag 1.
The colors correspond to the final clustering based the highest average silhouette width of 0.7 obtained with 3 clusters.}
\label{fig_new_pop}
\end{figure}
}

Consider the population of twenty states in the U.S.A.~between the years of 1900--1999.
This dataset, with dimension 20 and length $n=99$, was studied in \citet{Kalpakisetal2001, ZhangAn2018, ZhangChen2018} and is made available from \href{https://www.csee.umbc.edu/~kalpakis/TS-mining/ts-datasets.html}{https://www.csee.umbc.edu/~kalpakis/TS-mining/ts-datasets.html}.
\citet{Kalpakisetal2001} identified two clusters, where one set of states had an exponentially increasing trend whereas the second set of states had a stabilizing trend.
The states in the first cluster were
California, Colorado, Florida, Georgia, Maryland, North Carolina, South Carolina, Tennessee, Texas, Virginia and Washington.
The latter cluster consisted of Illinois, Massachusetts, Michigan, New Jersey, New York, Oklahoma, Pennsylvania, North Dakota and South Dakota.
The raw data was used to calculate the log growth rate for each state, that is, $\log (\text{P}_{t,j}) - \log(\text{P}_{t-1,j})$, where $\text{P}_{t,j}$ is the population of state $j$ at time $t$.
Each of the time series were normalized to have zero mean and unit standard deviation.
The results of clustering with $h=1$ are shown in Figure \ref{fig_new_pop}; the corresponding map of the U.S.A.~is displayed in Figure \ref{fig_pop_map}.
In this example the clustering was remarkably consistent when $h$  varied from 0 to 5.
This suggests that the stationary distributions differ between clusters.
The average silhouette score suggest three clusters in this case.
California, Colorado, Florida, Georgia, Maryland, Virginia, Washington, Illinois and Massachusetts are included in the red cluster and all but the last two have exponentially increasing trends.
North Carolina, South Carolina and Tennessee however were assigned to the green cluster.
Texas forms it's own separate cluster; the population growth data of Texas has different distribution compared to all the other states in this experiment.
The trend based clustering assignments of \citet{Kalpakisetal2001} are partially recovered and our method identifies the differences in the underlying (lagged) stationary distributions among the obtained clusters.

\ifthenelse{\boolean{withgraph}}
{
\begin{figure}[t]
\centering
\scalebox{0.9}{\input{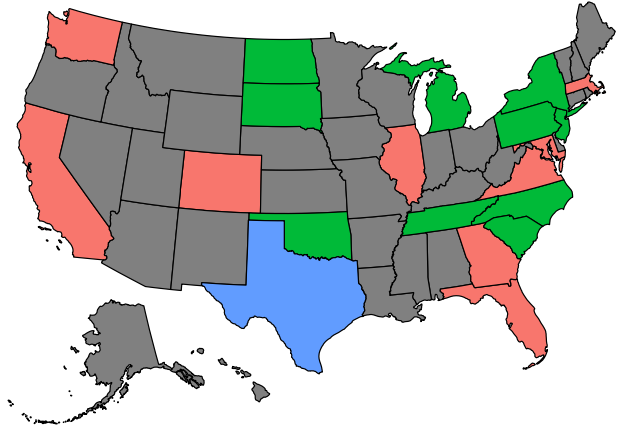}}
\caption{Clustering obtained from population data using energy distance with lag 1 on a map of the U.S.A.~in the years 1900--1999.}
\label{fig_pop_map}
\end{figure}
}

\section*{Acknowledgements}
The research of R. A. Davis was supported in part by NSF grant DMS 2015379 to Columbia University.
We acknowledge computing resources from Columbia University's Shared Research Computing Facility project, which is supported by NIH Research Facility Improvement Grant 1G20RR030893-01, and associated funds from the New York State Empire State Development, Division of Science Technology and Innovation (NYSTAR) Contract C090171, both awarded April 15, 2010.  We also thank the referees for their helpful remarks which led to a greatly improved exposition.

\newpage

\bibliographystyle{finaljtsa}
\bibliography{jtsabib}

\begin{thebibliography}{38}
\providecommand{\natexlab}[1]{#1}

\bibitem[{Aaronson \emph{et~al.}(1996)Aaronson, Burton, Dehling, Gilat, Hill
  and Weiss}]{AaronsonBurtonDehlingGilatHillWeiss1996}
Aaronson J, Burton R, Dehling H, Gilat D, Hill T, Weiss B. 1996.
\newblock Strong laws for {$L$}- and {$U$}-statistics.
\newblock \emph{Transactions of the American Mathematical Society},
  \textbf{348}(7):2845--2866.

\bibitem[{Aghabozorgi \emph{et~al.}(2015)Aghabozorgi, Shirkhorshidi and
  Wah}]{Aghabozorgi_etal2015}
Aghabozorgi S, Shirkhorshidi AS, Wah TY. 2015.
\newblock Time-series clustering--a decade review.
\newblock \emph{Information systems}, \textbf{53}:16--38.

\bibitem[{Batagelj(1988)}]{Batagelj1988}
Batagelj V. 1988.
\newblock Generalized Ward and Related Clustering Problems.
\newblock \emph{Classification and Related Methods of Data Analysis},
  \textbf{30}:67--74.

\bibitem[{Berndt and Clifford(1994)}]{BerndtJames1994}
Berndt DJ, Clifford J. 1994.
\newblock Using Dynamic Time Warping to Find Patterns in Time Series.
\newblock In \emph{Proceedings of the 3rd International Conference on Knowledge
  Discovery and Data Mining}, AAAIWS'94, 359–370. AAAI Press.

\bibitem[{Caiado \emph{et~al.}(2006)Caiado, Crato and
  Pe\'{n}a}]{Caiadoetal2006}
Caiado J, Crato N, Pe\'{n}a D. 2006.
\newblock A periodogram-based metric for time series classification.
\newblock \emph{Computational Statistics \& Data Analysis},
  \textbf{50}(10):2668--2684.

\bibitem[{Casado~de Lucas(2010)}]{Casado2010}
Casado~de Lucas D. 2010.
\newblock \emph{Classification techniques for time series and functional data}.
\newblock Ph.D. thesis, Universidad Carlos III de Madrid.

\bibitem[{Davis \emph{et~al.}(2018)Davis, Matsui, Mikosch and
  Wan}]{DavisMatsuiMikoschWan2018}
Davis RA, Matsui M, Mikosch T, Wan P. 2018.
\newblock Applications of distance correlation to time series.
\newblock \emph{Bernoulli}, \textbf{24}(4A):3087--3116.

\bibitem[{D\'{i}az and Vilar(2010)}]{DiazVilar2010}
D\'{i}az SP, Vilar JA. 2010.
\newblock Comparing several parametric and nonparametric approaches to time
  series clustering: a simulation study.
\newblock \emph{Journal of Classification}, \textbf{27}(3):333--362.

\bibitem[{Doukhan(1994)}]{Doukhan1994}
Doukhan P. 1994.
\newblock \emph{Mixing}, volume~85 of \emph{Lecture Notes in Statistics}.
\newblock New York: Springer-Verlag.

\bibitem[{Fokianos and Pitsillou(2018)}]{FokianosPitsillou2018}
Fokianos K, Pitsillou M. 2018.
\newblock Testing independence for multivariate time series via the
  auto-distance correlation matrix.
\newblock \emph{Biometrika}, \textbf{105}:337--352.

\bibitem[{Fu(2011)}]{Fu2011}
Fu TC. 2011.
\newblock A review on time series data mining.
\newblock \emph{Engineering Applications of Artificial Intelligence},
  \textbf{24}(1):164--181.

\bibitem[{Galeano and Pe\~{n}a(2000)}]{GaleanoPena2000}
Galeano P, Pe\~{n}a D. 2000.
\newblock Multivariate analysis in vector time series.
\newblock \emph{Resenhas do Instituto de Matem\'{a}tica e Estat\'{i}stica da
  Universidade de S\~{a}o Paulo}, \textbf{4}(4):383--403.

\bibitem[{Gavrilov \emph{et~al.}(2000)Gavrilov, Anguelov, Indyk and
  Motwani}]{Gavrilovetal2000}
Gavrilov M, Anguelov D, Indyk P, Motwani R. 2000.
\newblock Mining the stock market (extended abstract) which measure is best?
\newblock In \emph{Proceedings of the sixth ACM SIGKDD International Conference
  on Knowledge Discovery and Data Mining}, 487--496.

\bibitem[{Gretton \emph{et~al.}(2012)Gretton, Borgwardt, Rasch, Sch{\"o}lkopf
  and Smola}]{Gretton_etal2012}
Gretton A, Borgwardt KM, Rasch MJ, Sch{\"o}lkopf B, Smola A. 2012.
\newblock A kernel two-sample test.
\newblock \emph{The Journal of Machine Learning Research},
  \textbf{13}(1):723--773.

\bibitem[{Hong \emph{et~al.}(2017)Hong, Gu and
  Whitehouse}]{HongGuWhitehouse2017}
Hong D, Gu Q, Whitehouse K. 2017.
\newblock High-dimensional time series clustering via cross-predictability.
\newblock In \emph{Artificial Intelligence and Statistics}, 642--651. PMLR.

\bibitem[{Hubert and Arabie(1985)}]{HubertArabie1985}
Hubert L, Arabie P. 1985.
\newblock Comparing partitions.
\newblock \emph{Journal of Classification}, \textbf{2}(1):193--218.

\bibitem[{James \emph{et~al.}(2013)James, Witten, Hastie and
  Tibshirani}]{Jamesetal2013}
James G, Witten D, Hastie T, Tibshirani R. 2013.
\newblock \emph{An Introduction to Statistical Learning}.
\newblock New York: Springer.

\bibitem[{Kakizawa \emph{et~al.}(1998)Kakizawa, Shumway and
  Taniguchi}]{Kakizawa_etall1998}
Kakizawa Y, Shumway RH, Taniguchi M. 1998.
\newblock Discrimination and clustering for multivariate time series.
\newblock \emph{Journal of the American Statistical Association},
  \textbf{93}:328--340.

\bibitem[{Kalpakis \emph{et~al.}(2001)Kalpakis, Gada and
  Puttagunta}]{Kalpakisetal2001}
Kalpakis K, Gada D, Puttagunta V. 2001.
\newblock Distance measures for effective clustering of ARIMA time-series.
\newblock In \emph{Proceedings 2001 IEEE International Conference on Data
  Mining}, 273--280. IEEE.

\bibitem[{Kaufman and Rousseeuw(2009)}]{KaufmanRousseeuw2009}
Kaufman L, Rousseeuw PJ. 2009.
\newblock \emph{Finding Groups in Data: An Introduction to Cluster Analysis}.
\newblock New York: Wiley.

\bibitem[{Krengel(1985)}]{Krengel2011}
Krengel U. 1985.
\newblock \emph{Ergodic Theorems}.
\newblock Berlin: Walter de Gruyter \& Co.

\bibitem[{Liao(2005)}]{Liao2005}
Liao TW. 2005.
\newblock Clustering of time series data—a survey.
\newblock \emph{Pattern Recognition}, \textbf{38}(11):1857--1874.

\bibitem[{Maharaj(2000)}]{Maharaj2000}
Maharaj EA. 2000.
\newblock Cluster of Time Series.
\newblock \emph{Journal of Classification}, \textbf{17}(2):297--314.

\bibitem[{Maharaj \emph{et~al.}(2019)Maharaj, D'Urso and
  Caido}]{MaharajD'UrsoCaido2019}
Maharaj EA, D'Urso P, Caido J. 2019.
\newblock \emph{Time Series Clustering and Classification}.
\newblock Computer Science and Data Analysis. Australia: CRC Press.

\bibitem[{Montero and Vilar(2014)}]{MonteroVilar2015}
Montero P, Vilar JA. 2014.
\newblock TSclust: An {$\mathtt{R}$} Package for Time Series Clustering.
\newblock \emph{Journal of Statistical Software}, \textbf{62}(1):1–43.

\bibitem[{Murtagh and Legendre(2014)}]{MurtaghLegendre2014}
Murtagh F, Legendre P. 2014.
\newblock Ward's Hierarchical Agglomerative Clustering Method: Which Algorithms
  Implement Ward's Criterion?
\newblock \emph{Journal of Classification}, \textbf{31}(3):274--295.

\bibitem[{Piccolo(1990)}]{Piccolo1990}
Piccolo D. 1990.
\newblock A distance measure for classifying ARIMA models.
\newblock \emph{Journal of Time Series Analysis}, \textbf{11}(2):153--164.

\bibitem[{Rand(1971)}]{Rand1971}
Rand WM. 1971.
\newblock Objective criteria for the evaluation of clustering methods.
\newblock \emph{Journal of the American Statistical Association},
  \textbf{66}(336):846--850.

\bibitem[{Savvides \emph{et~al.}(2008)Savvides, Promponas and
  Fokianos}]{Savvidesetal2008}
Savvides A, Promponas VJ, Fokianos K. 2008.
\newblock Clustering of biological time series by cepstral coefficients based
  distances.
\newblock \emph{Pattern Recognition}, \textbf{41}(7):2398--2412.

\bibitem[{Shumway(1982)}]{Shumway1982}
Shumway RH. 1982.
\newblock Discriminant analysis for time series.
\newblock In \emph{Classification, pattern recognition and reduction of
  dimensionality}, volume~2 of \emph{Handbook of Statistics}, 1--46. Amsterdam:
  North-Holland.

\bibitem[{Stout(1974)}]{Stout1974}
Stout WF. 1974.
\newblock \emph{Almost sure convergence}.
\newblock New York-London: Academic Press.

\bibitem[{Sz\'{e}kely and Rizzo(2013)}]{SzekelyRizzo2013}
Sz\'{e}kely GJ, Rizzo ML. 2013.
\newblock Energy statistics: A class of statistics based on distances.
\newblock \emph{Journal of Statistical Planning and Inference},
  \textbf{143}(8):1249--1272.

\bibitem[{Sz\'{e}kely \emph{et~al.}(2007)Sz\'{e}kely, Rizzo and
  Bakirov}]{SzekelyRizzoBakirov2007}
Sz\'{e}kely GJ, Rizzo ML, Bakirov NK. 2007.
\newblock Measuring and testing dependence by correlation of distances.
\newblock \emph{The Annals of Statistics}, \textbf{35}(6):2769--2794.

\bibitem[{Tan \emph{et~al.}(2006)Tan, Steinbach and Kumar}]{Tanetal2006}
Tan PN, Steinbach M, Kumar V. 2006.
\newblock \emph{Data Mining Introduction}.
\newblock Boston: Pearson Addison Wesley.

\bibitem[{Taniguchi and Kakizawa(2000)}]{TaniguchiandKakizawa2000}
Taniguchi M, Kakizawa Y. 2000.
\newblock \emph{Asymptotic Theory of Statistical Inference for Time Series}.
\newblock New York: Springer.

\bibitem[{Xu and Tian(2015)}]{XuTian2015}
Xu D, Tian Y. 2015.
\newblock A comprehensive survey of clustering algorithms.
\newblock \emph{Annals of Data Science}, \textbf{2}(2):165--193.

\bibitem[{Zhang and An(2018)}]{ZhangAn2018}
Zhang B, An B. 2018.
\newblock Clustering time series based on dependence structure.
\newblock \emph{PLoS ONE}, \textbf{13}(11):1--22.

\bibitem[{Zhang and Chen(2018)}]{ZhangChen2018}
Zhang B, Chen R. 2018.
\newblock Nonlinear Time Series Clustering Based on Kolmogorov-Smirnov 2D
  Statistic.
\newblock \emph{Journal of Classification}, \textbf{35}(3):394--421.

\end{thebibliography}

\appendix

\section{Appendices}
\renewcommand{\thesubsection}{\thesection.\arabic{subsection}}

\subsection{Proof of Lemma \ref{lemma.denergy}}
From Lemma 1 in \citet{SzekelyRizzoBakirov2007}, for $x\in\R^p$
\begin{align}
\label{eq_denergyprop}
\int_{\R^p} \frac{1-\cos \innerproduct{s}{x}}{|s|^{p+1}c_p}ds = |x|.
\end{align}
Let $(\dot Y,\dot Z)$ be an independent copy of $(Y,Z)$.
For $s\in\R^p$,
\begin{align*}
	\big| \varphi_Y(s) - \varphi_Z(s) \big|^2 &= |\varphi_Y(s)|^2 + |\varphi_Z(s)|^2 - \varphi_Y(s)\overline{\varphi_Z(s)} - \overline{\varphi_Y(s)}\varphi_Z(s)\\
	&= \E e^{i \innerproduct{s}{Y - \dot Y} } + \E e^{i \innerproduct{s}{Z - \dot Z} } - \E e^{i \innerproduct{s}{Y - \dot Z} } - \E e^{i \innerproduct{s}{\dot Z - Y} }\\
	&= \E (\cos  \innerproduct{s}{Y - \dot Y} ) + \E (\cos  \innerproduct{s}{Z - \dot Z} ) - 2\E ( \cos \innerproduct{s}{Y- \dot Z} ) \\
	&= 2 \E ( 1 - \cos \innerproduct{s}{Y - \dot Z} ) - \E (1-\cos  \innerproduct{s}{Y - \dot Y} ) - \E (1- \cos  \innerproduct{s}{Z - \dot Z} )\,.
\end{align*}
Since $\E [|Y| + |Z|] < \infty$, we can apply Fubini's theorem and \eqref{eq_denergyprop} to deduce $d_E(Y,Z) < \infty$ and obtain \eqref{eq_denergy.exp}.

\subsection{Proof of Theorem \ref{thm_cons_direct}}
We follow the steps from the proofs of Theorem 3.1 in \citet{DavisMatsuiMikoschWan2018} and Theorem 2 in \citet{SzekelyRizzoBakirov2007}.
Accordingly, for $\delta>0$ we set
\begin{align}
	\label{KDelta}
	K_{\delta} = \{ s \in \R^p: \delta \leq |s| \leq 1/\delta\}.
\end{align}
The processes $\hat\varphi_Y$ and $\hat\varphi_Z$ are sample means of i.i.d bounded processes.
By the ergodic theorem (Theorem 3.5.7 in \citet{Stout1974}) $\hat\varphi_Y \gas \varphi_Y$ and $\hat\varphi_Z \gas \varphi_Z$ on $\mathcal{C}(K_{\delta})$, the space of continuous functions on $K_{\delta}$; see \citet{Krengel2011}.
So,
\begin{align*}
	\int_{K_{\delta}} \big|\hat\varphi_Y(s)-\hat\varphi_Z(s)\big|^2 d\mu(s) \gas \int_{K_{\delta}} \big|\varphi_Y(s)-\varphi_Z(s)\big|^2 d\mu(s).
\end{align*}
Thus it suffices to show
\begin{align}
	\label{eq.cons.etp}
	\lim_{\delta\downarrow 0} \limsup_{n\rightarrow\infty} \int_{K_{\delta}^c} \big| \hat\varphi_Y(s) -\hat\varphi_Z(s) \big|^2 d\mu(s) \gas 0.
\end{align}
First, since $\big| \hat\varphi_Y(s) -\hat\varphi_Z(s) \big|^2 \leq 4$ we have almost surely
\begin{align*}
	\lim_{\delta\downarrow 0} \limsup_{n\rightarrow\infty} \int_{|s|>1/\delta} \big| \hat\varphi_Y(s) -\hat\varphi_Z(s) \big|^2 d\mu(s) \leq 4 \lim_{\delta\downarrow 0} \int_{|s|>1/\delta} d\mu(s)  = 0.
\end{align*}
Fix $\delta >0$.
From the proof of Lemma \ref{lemma.denergy}, we rewrite
\begin{align*}
\int_{|s|<\delta} |\hat \varphi_Y(s) - \hat \varphi_Z(s)|^2 d\mu(s)
&=  \int_{|s|<\delta} \bigg[ \frac{2}{n^2} \sum_{j,k=1}^{n} (1 - \cos  \innerproduct{s}{Y_j - Z_k} ) \bigg] d\mu(s) \\
&- \int_{|s|<\delta} \bigg[  \frac{1}{n^2} \sum_{j,k=1}^{n} (1-\cos  \innerproduct{s}{Y_j - Y_k} ) \bigg] d\mu(s) \\
	&- \int_{|s|<\delta} \bigg[  \frac{1}{n^2} \sum_{j,k=1}^{n}(1-\cos  \innerproduct{s}{Z_j - Z_k} ) \bigg] d\mu(s).
\end{align*}
Let $g(x):= \int_{|s|<x} (1-\cos (s_1)) \frac{ds}{c_ps^2}$, where $s_1$ is the first component of $s\in\R^p$.
Due to a change of variables,
\begin{align*}
\int_{|s|<\delta} |\hat \varphi_Y(s) - \hat \varphi_Z(s)|^2 d\mu(s)
	&= \frac{2}{n^2} \sum_{j,k=1}^{n} g(|Y_j-Z_k|\delta)|Y_j-Z_k| \\
	&- \frac{1}{n^2} \sum_{j,k=1}^{n} g(|Y_j-Y_k|\delta)|Y_j-Y_k| \\
	&- \frac{1}{n^2} \sum_{j,k=1}^{n} g(|Z_j-Z_k|\delta)|Z_j-Z_k|.
\end{align*}
Applying the ergodic theorem for U-statistics according to \citet{AaronsonBurtonDehlingGilatHillWeiss1996}, as $n \rightarrow\infty$
\begin{align*}
	\int_{|s|<\delta} |\hat \varphi_Y(s) - \hat \varphi_Z(s)|^2 d\mu(s) \gas
	&2\E[ g(|Y-Z|\delta)|Y-Z| ]  - \E[ g(|Y-\dot Y|\delta)|Y-\dot Y| ]  \\
	&- \E[ g(|Z-\dot Z|\delta)|Z-\dot Z| ].
\end{align*}
Using the continuity of $g(\cdot)$ at zero we see $\lim_{\delta \downarrow 0} \E[ g(|Y-Z|\delta)|Y-Z| ] = 0$ via dominated convergence. Similarly for the remaining two terms.
It then follows that almost surely,
\begin{align*}
	\lim_{\delta\downarrow 0} \limsup_{n\rightarrow\infty} \int_{|s|<\delta} \big| \hat\varphi_Y(s) -\hat\varphi_Z(s) \big|^2 d\mu(s) = 0.
\end{align*}
Hence \eqref{eq.cons.etp} holds, which concludes the proof.

\subsection{Proof of Theorem \ref{thm_asymptotic_direct}}

In the proof below, the symbol $c$ will denote a positive constant, whose value might
change from line to line but it is not of particular interest.
For $s\in\R^p$, due to stationarity of $\{Y_t\}$,
\begin{align*}
	\E \Big[ \Big| \frac{1}{n} \sum_{j=1}^n \big( e^{i \innerproduct{s}{Y_j}} - \varphi_Y(s) \big) \Big|^2 \Big]
	&= \frac{1}{n} \sum_{h=1-n}^{n-1} \big(1-|h|/n \big) \RE \big[ \, \Cov \big(  e^{i \innerproduct{s}{Y_0}} - \varphi_Y(s) , e^{i \innerproduct{s}{Y_h}} - \varphi_Y(s) \big) \big].
\end{align*}
An application of Theorem 3(a) of Section 1.2.2 in \citet{Doukhan1994} yields
\begin{align*}
	\big| \RE \big[ \, \Cov \big(  e^{i \innerproduct{s}{Y_0}} - \varphi_Y(s) , e^{i \innerproduct{s}{Y_h} } - \varphi_Y(s) \big) \big] \big|
	&\leq c \alpha_h^{1/r} \big( \E \big[ \big| e^{i \innerproduct{s}{Y}} - \varphi_Y(s) \big|^u \big] \big)^{2/u} \\
	&\leq c \alpha_h^{1/r} \big( \E \big[ \big| e^{i \innerproduct{s}{Y}} - \varphi_Y(s) \big|^2 \big] \big)^{2/u}.
\end{align*}
Then, for $\alpha \in (0,2]$
\begin{align*}
	\E \big[ \big| e^{i \innerproduct{s}{Y}} - \varphi_Y(s) \big|^2 \big] = 1 - | \varphi_Y(s) |^2 \leq \E [ 1 \wedge | \innerproduct{s}{Y - \dot Y} |^{\alpha} ] \leq c( 1\wedge |s|^{\alpha}).
\end{align*}
Due to the summability assumption $\sum_h \alpha_h^{1/r} < \infty$,
\begin{align*}
	n\E \big[ \big| \hat \varphi_Y(s) - \varphi_Y(s) \big|^2 \big]
	&\leq c (1\wedge |s|^{2\alpha/u}) \sum_{h=1-n}^{n-1} (1-|h|/n)\alpha_h^{1/r}\\
	&\leq c(1\wedge |s|^{2\alpha/u}).
\end{align*}
Similarly, $n \E \big[ \big| \hat \varphi_Z(s) - \varphi_Z(s) \big|^2 \big] \leq c(1\wedge |s|^{2\alpha/u})$.
It then follows that
\begin{align}
    \label{eq.chfnbound}
	\E[ \hat G(s)^2] \leq c ( 1 \wedge |s|^{2\alpha/u}),
\end{align}
where $\hat G(s) := \sqrt{n} \big( (\hat\varphi_Y(s) - \varphi_Y(s)) - (\hat\varphi_Z(s) - \varphi_Z(s)) \big)$.  The proof of the theorem will rely on  Lemma A.1(2) of \citet{DavisMatsuiMikoschWan2018} stated below.
\begin{lemma}
		\label{lemma.ecfn}
    Assume that $\sum_h \alpha_h^{1/r} <\infty$ for some $r>1$ and set $u=2r/(r-1)$.
    If the moment conditions \eqref{eq.same.moment} are satisfied with $u/2 < \alpha \leq u$, then $\sqrt{n} (\hat \varphi_{Y,Z}-\varphi_{Y,Z}) \gid G_{Y,Z}$ on compact sets $K\subset \R^{2p}$ for some complex-valued mean-zero Gaussian field $G_{Y,Z}$ with covariance structure
		\begin{equation*}
			\Cov (G_{Y,Z}(s), G_{Y,Z}(t)) = \sum_{h \in \Z} \Cov (e^{i \langle s, Y_0 \rangle+ i \langle t, Z_0 \rangle)}, e^{i \langle s, Y_h \rangle+ i \langle t, Z_h \rangle)}).
		\end{equation*}
\end{lemma}
Due to Lemma \ref{lemma.ecfn}, on compact sets $\sqrt{n} (\hat \varphi_{Y,Z}-\varphi_{Y,Z}) \gid G_{Y,Z}$.
But $\hat G(s) = \sqrt{n} (\hat \varphi_{Y,Z}(s,0) - \varphi_{Y,Z}(0,s))$.
So on the compact set $K_{\delta}$ defined in \eqref{KDelta} for some $\delta>0$, we obtain $\hat G \gid G$,
where $G$ is a complex-valued mean-zero Gaussian process.
The covariance structure is then given by
\begin{align*}
    \Cov (G(s), G(t)) &= \Cov (G_{Y,Z}(s,0) - G_{Y,Z}(0,s), G_{Y,Z}(t,0) - G_{Y,Z}(0,t)) \\
    &= \Cov (G_{Y,Z}(s,0), G_{Y,Z}(t,0)) -
    \Cov (G_{Y,Z}(s,0), G_{Y,Z}(0,t)) \\&-
    \Cov (G_{Y,Z}(0,s), G_{Y,Z}(t,0) +
    \Cov (G_{Y,Z}(0,s), G_{Y,Z}(0,t)) \\
    &= \sum_{h \in \Z} \big( \Cov \big(e^{i \langle s, X_0 \rangle}, e^{i \langle t, X_h \rangle}\big) -
    \Cov \big(e^{i \langle s, X_0 \rangle}, e^{i \langle t, Y_h \rangle}\big)
		\\&-
    \Cov \big(e^{i \langle s, Y_0 \rangle}, e^{i \langle t, X_h \rangle}\big)  +
    \Cov \big(e^{i \langle s, Y_0 \rangle}, e^{i \langle t, Y_h \rangle}\big) \big) \\
    &= \sum_{h \in \Z} \Cov \big(e^{i \langle s, X_0 \rangle} - e^{i \langle s, Y_0 \rangle}, e^{i \langle s, Y_0 \rangle} - e^{i \langle s, Y_h \rangle}\big).
\end{align*}

\paragraph{Proof of Theorem \ref{thm_asymptotic_direct}(i).}
If $Y \eqd Z$, then $\varphi_Y = \varphi_Z$ so that $\sqrt{n} \big( \hat\varphi_Y - \hat\varphi_Z \big) \gid G$ on $K_{\delta}$.
By the continuous mapping theorem,
\begin{align*}
	\int_{K_{\delta}} \big| \sqrt{n} \big( \hat\varphi_Y(s) - \hat\varphi_Z(s) \big) \big|^2 d\mu(s) \gid \int_{K_{\delta}} \big| G(s) \big|^2 d\mu(s).
\end{align*}
Let $\varepsilon>0$.
From Markov's inequality, dominated convergence and \eqref{eq.chfnbound},
\begin{align*}
	\lim_{\delta\downarrow 0} & \limsup_{n\rightarrow\infty} \mathbb{P} \bigg( \int_{K_{\delta}^c} \Big| \sqrt n \Big( \hat \varphi_Y(s) - \hat \varphi_Z(s) \Big)  \Big|^2 d\mu(s) > \varepsilon \bigg)\\
	& \leq \varepsilon^{-1} \lim_{\delta\downarrow 0} \limsup_{n\rightarrow\infty} \int_{K_{\delta}^c} n \E \Big[ \Big| \hat \varphi_Y(s) - \hat \varphi_Z(s) \Big|^2 \Big] d\mu(s)\\
	& \leq c \varepsilon^{-1}  \lim_{\delta\downarrow 0} \int_{K_{\delta}^c} ( 1 \wedge |s|^{2\alpha/u}) d\mu(s) = 0.
\end{align*}
\qed

\paragraph{Proof of Theorem \ref{thm_asymptotic_direct}(ii).}
Now $Y$ and $Z$ have different distributions.
Note that
\begin{align*}
	\sqrt{n} \big( \big| \hat \varphi_Y(s) - \hat \varphi_Z(s) \big|^2 - \big| \varphi_Y(s) - \varphi_Z(s) \big|^2 \big)
	&= \hat G(s) (\hat\varphi_Y(-s) - \hat\varphi_Z(-s))
	+ \hat G(-s) (\varphi_Y(s) - \varphi_Z(s)).
\end{align*}
From the almost sure convergence of $\hat\varphi_Y(\cdot) - \hat\varphi_Z(\cdot)$ on compact sets and the continuous mapping theorem,
\begin{align*}
	\sqrt{n} \int_{K_{\delta}} \big( \big| \hat \varphi_Y(s) - \hat \varphi_Z(s) \big|^2 - \big| \varphi_Y(s) - \varphi_Z(s) \big|^2 \big) d\mu(s) \gid \int_{K_{\delta}} G'(s) d\mu(s),
\end{align*}
where $G'(s) = 2\RE[(\varphi_Y(s)-\varphi_Z(s))\cdot \overline{G(s)}]$.
Note that for complex numbers $z_1,z_2$ we have
\[
\big| |z_1|^2 - |z_2|^2 \big| \leq \big| \RE\, (z_1 - z_2)(\overline{z_1} + \overline{z_2}) \big|.
\]
Using this for $z_1 = \hat\varphi_Y(s) - \hat\varphi_Z(s)$ and $z_2 = \varphi_Y(s) - \varphi_Z(s)$,
\[
\big| | \hat\varphi_Y(s) - \hat\varphi_Z(s) |^2 -  | \varphi_Y(s) - \varphi_Z(s) |^2 \big| \leq c  \big| ( \hat\varphi_Y(s) - \hat\varphi_Z(s) ) -  ( \varphi_Y(s) - \varphi_Z(s))  \big|.
\]
Therefore, by Markov's inequality, Jensen's inequality, the dominated convergence theorem and \eqref{eq.chfnbound}, for any given $\varepsilon > 0$,
\begin{align*}
	\lim_{\delta\downarrow 0} & \limsup_{n\rightarrow\infty} \mathbb{P} \bigg( \int_{K_{\delta}^c} \sqrt{n} \Big| \big| \hat \varphi_Y(s) - \hat \varphi_Z(s) \big|^2 - \big| \varphi_Y(s) - \varphi_Z(s) \big|^2 \Big| d\mu(s) > \varepsilon \bigg)\\
	& \leq c \lim_{\delta\downarrow 0} \limsup_{n\rightarrow\infty} \int_{K_{\delta}^c} \sqrt{n} \E \big| ( \hat\varphi_Y(s) - \hat\varphi_Z(s) ) -  ( \varphi_Y(s) - \varphi_Z(s) ) | d\mu(s)\\
	& \leq c \lim_{\delta\downarrow 0} \limsup_{n\rightarrow\infty} \int_{K_{\delta}^c} (\E \hat G(s)^2)^{1/2} d\mu(s)\\
	& \leq c \lim_{\delta\downarrow 0} \int_{K_{\delta}^c} ( 1 \wedge |s|^{2\alpha/u}) d\mu(s) = 0.
\end{align*}

\qed


\end{document}